\documentclass[12pt]{iopart}
\usepackage{epsfig}
\usepackage{longtable}
\usepackage{float}
\usepackage{rotating}
\usepackage{amssymb}
\usepackage{latexsym}

\newcommand{\Au} {\mbox{$\mathrm{Au}+\mathrm{Au}$}}
\newcommand{\Cu} {\mbox{$\mathrm{Cu}+\mathrm{Cu}$}}
\newcommand{\pp} {\mbox{$\mathrm{p}+\mathrm{p}$}}
\newcommand{\piz} {\mbox{$\pi^0$}}

\begin{document}

\title[PHENIX High-$p_T$ Hadron-hadron and Photon-hadron Azimuthal Correlations]{PHENIX Measurement of High-$p_T$ Hadron-hadron and Photon-hadron Azimuthal Correlations}

\author{J Jin for the PHENIX(\footnote[1]{For the full list of PHENIX authors and
acknowledgements, see Appendix 'Collaborations' of this volume})
Collaboration}

\address{Columbia University, New York, NY 10027, USA} \ead{jiamin@phys.columbia.edu}
\begin{abstract}

High-$p_T$ hadron-hadron correlations have been measured with the
PHENIX experiment in $\Cu$ and $\pp$ collisions at
$\sqrt{s_{NN}}=200$~GeV. A comparison of the jet widths and yields
between the two colliding systems allows us to study the medium
effect on jets. We also present a first measurement of direct
photon-hadron correlations in $\Au$ and $\pp$ collisions. We find
that the near-side yields are consistent with zero in both
systems. By comparing the jet yields on the away side, we observe
a suggestion of the expected suppression of hadrons associated
with photons in $\Au$ collisions.

\end{abstract}

%\submitto{\JP
%\maketitle

\section{Introduction}

The method of high-$p_T$ two-particle azimuthal correlations is a
unique probe of the hot, dense medium created in heavy-ion
collisions at RHIC. Early RHIC results on hadron-hadron
correlations indicate a strong modification of the away-side jet
shape and yield by the medium~\cite{paul}. These modifications
provide valuable constraints on the properties of the hot, dense
medium. However, the physics interpretations of the away-side
modification are complicated as the trigger hadrons mostly come
from the surface of the medium. Direct photons, due to their weak
coupling with the medium, provide a cleaner calibration of the
energy and direction of the away-side jets. Thus, direct
photon-hadron correlations can provide less biased and
quantitative measurements of the away-side modifications.

At RHIC energies, identification of direct photons is difficult
due to the large number of background photons from hadronic
decays, mostly from $\piz$ decays. Therefore the extraction of the
direct photon-hadron per-trigger yields relies on a statistical
subtraction of the decay photon-hadron per-trigger yields from the
inclusive photon-hadron per-trigger yields.

A two-particle correlation function ($C(\Delta\phi)$) as measured
in the PHENIX central spectrometer arms is constructed as
\begin{equation}
CF(\Delta\phi)\sim\frac{dN_{real}/d\Delta\phi}{dN_{mix}/d\Delta\phi}=Jet(\Delta\phi)+Bkdg(\Delta\phi)
\end{equation}
where $dN_{real}/d\Delta\phi$ is the same-event pair distribution
and $dN_{mix}/d\Delta\phi$ is the mixed-event pair distribution.
The mixed-event distribution is used to correct for the
non-uniform PHENIX pair acceptance. The correlation function (CF)
can be decomposed into a jet function $J(\Delta\phi)$ and a
underlying flow modulated background term. After subtracting the
background, we correct the remaining $J(\Delta\phi)$ by the single
particle efficiency and the PHENIX acceptance, then normalize it
by the number of triggers, thus obtaining the per-trigger
yield~\cite{refjjia}.

\section{High-$p_T$ hadron-hadron correlations}

The per-trigger yield distributions are fitted with a double
Gaussian function to extract the Gaussian width for both peaks
($\Delta\phi=0/\pi$ for near-side/away-side). The jet yield is
integrated over a $\Delta\pi$ region of $\pi$ around each peak.
Two other useful jet variables are defined as
\begin{equation}
p_{out}=p_{T,asso}\cdot\sin(\Delta\phi),\quad
x_{E}=\frac{p_{T,asso}}{p_{T,trig}}\cdot\cos(\Delta\phi)
\end{equation}
$p_{out}$ is the transverse momentum component of the associated
particle perpendicular to the trigger, %its distribution relates to
%the large angle hard radiation.
$x_{E}$ measures the relative associated particle $p_T$ to trigger
$p_T$ along the trigger direction.
% it is connected to the patonic $p_T$ spectrum in $\pp$.

\begin{figure}[ht]
\includegraphics[width=0.55\textwidth]{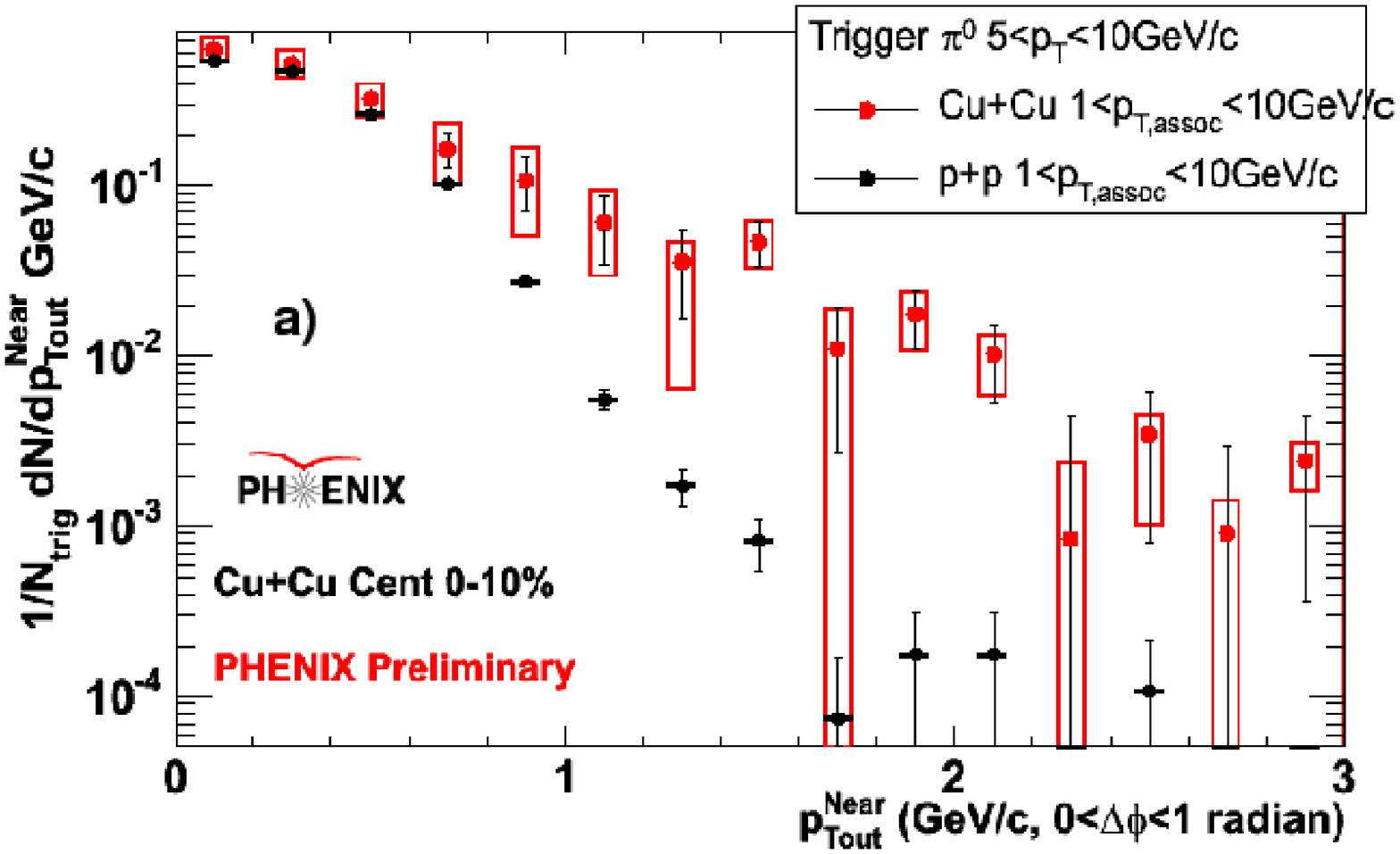}\hfill\includegraphics[width=0.4\textwidth]{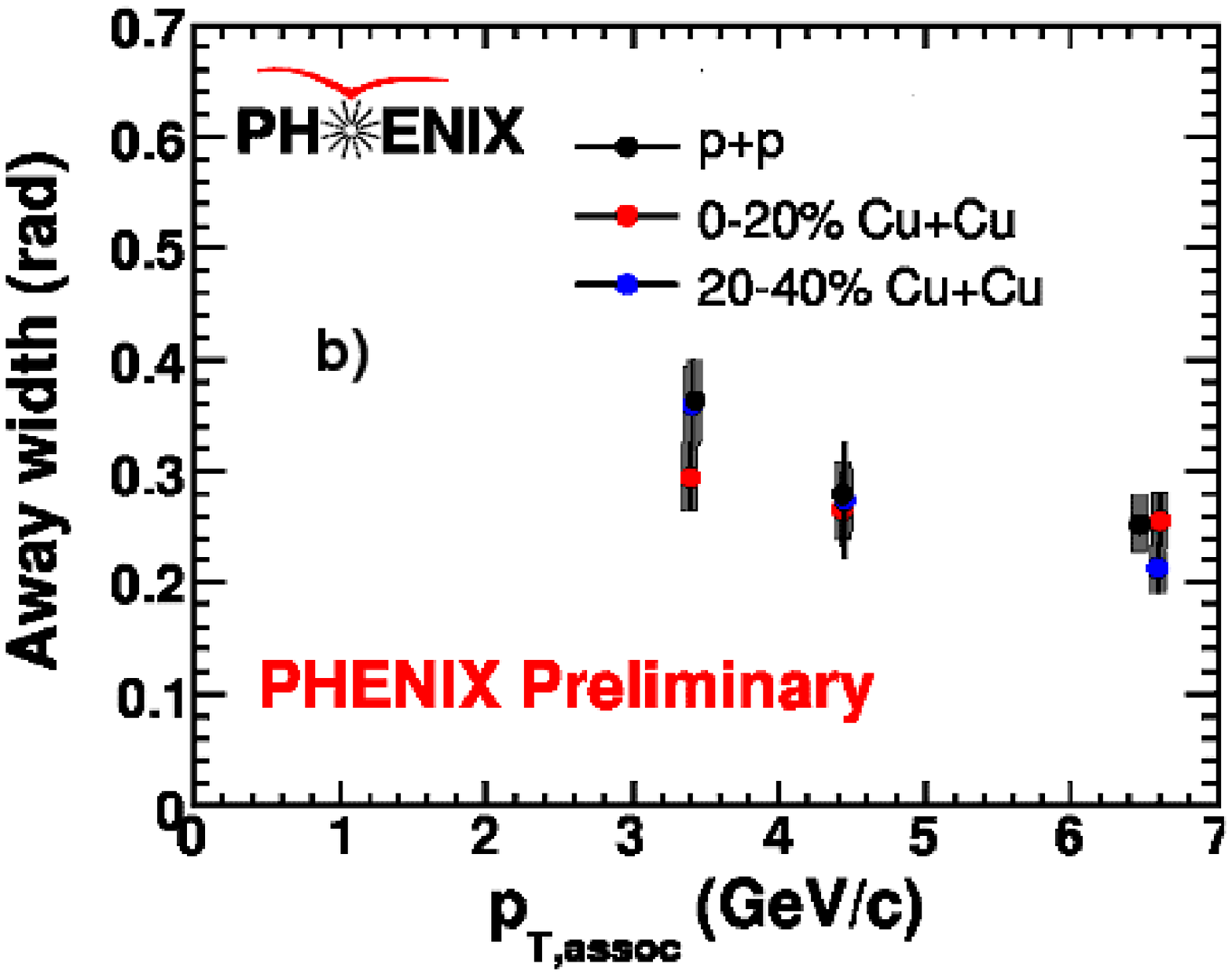}
\caption{(a) Near-side $p_{out}$ distribution, $\pp$ (black) and
$\Cu$ (red) and (b) Away side Gaussian width as a function of
associated $p_T$, $\pp$ (black), $\Cu$ $0-20\%$ (red) and $\Cu$
$20-40\%$ (blue) }\label{fig:nearside}
\end{figure}

%\begin{figure}
%\begin{minipage}{0.45\textwidth}
%\hfil\epsfig{file=fig/nearsidewidth.eps,width=0.9\linewidth}
%\caption{Near-side Gaussian width as a function of associated
%$p_T$, $\pp$ (black), $\Cu$ $0-20\%$ (red) and $\Cu$ $20-40\%$
%(blue).} \label{fig:nearside}
%\end{minipage}
%\hfill
%\begin{minipage}{0.45\textwidth}
%\hfil\epsfig{file=fig/pout.eps,width=0.9\linewidth} \caption{Near
%side $p_{out}$ distribution, $\pp$ (black) and $\Cu$ (red).}
%\label{fig:pout}
%\end{minipage}
%\end{figure}

Fig.~\ref{fig:nearside}(a) shows the near-side $p_{out}$
distribution. Near-side Gaussian widths are also extracted and no
significant difference is seen between $\Cu$ and $\pp$. However,
the $p_{out}$ distribution is a more sensitive quantity to study
medium-induced jet modification, especially at the large $p_{out}$
region. On Fig.~\ref{fig:nearside}(a), we see an enhancement in
$\Cu$ compared with $\pp$ in the tail region. It indicates that
the near-side jets are modified by the medium through additional
radiation with components transverse to the jet direction.
Fig.~\ref{fig:nearside}(b) shows the away-side Gaussian widths, we
do not see a width broadening from $\pp$ to $\Cu$. We are working
on the away-side $p_{out}$ distribution to better study the
away-side shape modification.

\begin{figure}[ht]
\includegraphics[width=0.45\textwidth]{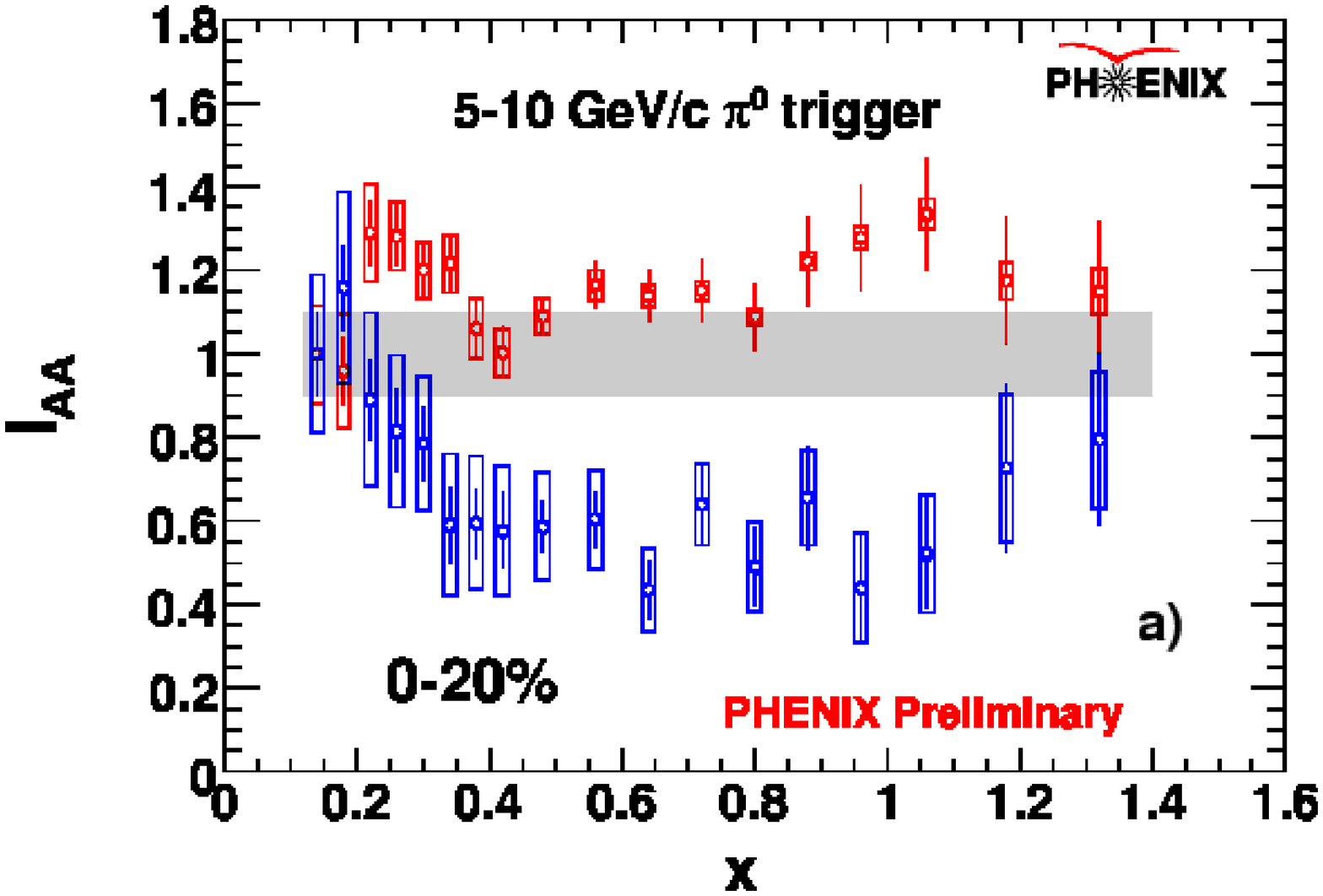}\hfil\includegraphics[width=0.4\textwidth]{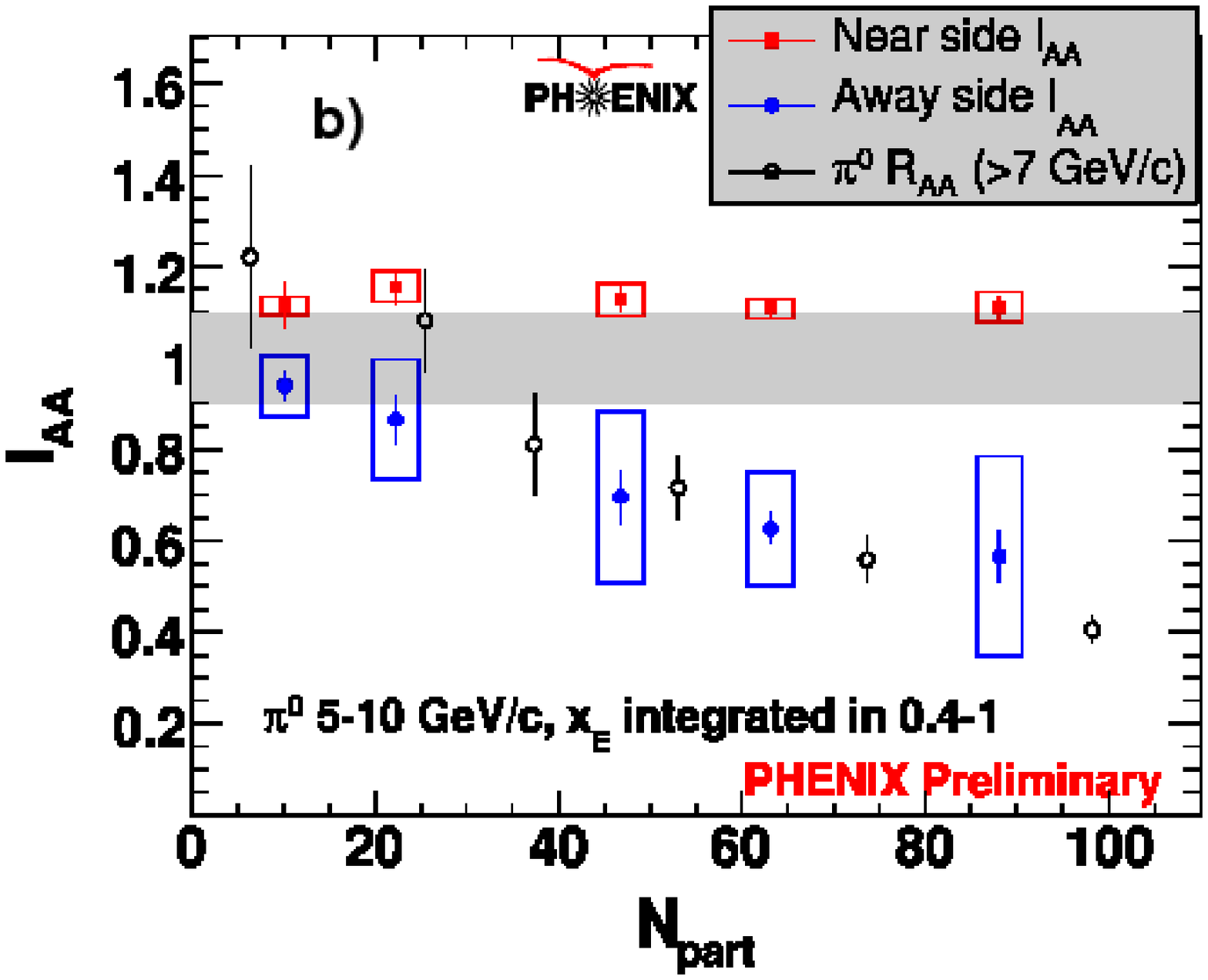}
\caption{(a) $I_{AA}$ as a function of associated $p_T$, near side
(red) and away side (blue). and (b) Integrated $I_{AA}$ and
$R_{AA}$ as a function of $N_{part}$, near side (red), away side
(blue) and $R_{AA}$ (black).}\label{fig:iaa}
\end{figure}

PHENIX measures jet yield modification via $I_{AA}$, which is the
yield ratio of $\Cu$ to $\pp$. $I_{AA}=1$ indicates no suppression
in $\Cu$ with respect to $\pp$ whereas the lower the $I_{AA}$, the
stronger the suppression. Fig.~\ref{fig:iaa}(a) shows $I_{AA}$ of
most central (0-20\%) $\Cu$ collisions. The near-side $I_{AA}$ is
close to unity and the away-side $I_{AA}$ shows substantial
suppression in the central $\Cu$ collisions. Fig.~\ref{fig:iaa}(b)
shows the $I_{AA}$ integrated over $x_E=0.4-1$ as a function of
$N_{part}$. The near-side $I_{AA}$ is consistent with unity within
error bars, whereas the away-side $I_{AA}$ shows a decreasing
trend from peripheral to central collisions. On the same plot is a
comparison to the nuclear modification factor $R_{AA}$ of
high-$p_T$ $\piz$s, the observed $I_{AA}$ is similar to the
nuclear modification factor.

\section{High-$p_T$ direct photon-hadron correlations}

While dijet measurements suffer from trigger bias and possible
trigger surface bias, an ideal probe for studying the jet
modification in medium is the use of direct photon-hadron
correlations~\cite{xinnianwang}. This measurement is aided by the
fact that PHENIX has observed a factor of $\sim$2 excess of
photons above the hadronic decay background at $p_T>5GeV/c$ in the
most central $\Au$ collisions, which is consistent with direct
photon production as calculated by pQCD~\cite{justin}. We employ a
statistical subtraction method to extract direct photon-hadron
correlations and per-trigger yields. First, the inclusive
photon-hadron per-trigger yields ($Y_{incl-h}$) are constructed by
subtracting the background term (using the measured inclusive
photon $v_2$) from the inclusive photon-hadron CF. Then, starting
from $\piz$-hadron pairs, we construct the decay photon-hadron CF.
This is done by performing a pair by pair weighted sum to
convolute the contributions from feeddown from $\piz$ decays. The
weights applied are derived from the $\piz$ decay kinematics. From
the decay photon-hadron CF, we subtract the background term using
the decay photon-hadron $v_2$ derived from the measured $\piz$
$v_2$ in order to finally get the decay photon-hadron per-trigger
yields. The inclusive photon and $\piz$ $v_2$'s that we used were
measured with the standard PHENIX single particle
method~\cite{v2}. Once both per-trigger yields are obtained, the
direct photon-hadron per-trigger yield ($Y_{dir-h}$) is found by:
\begin{equation}
Y_{dir-h}=\frac{1}{R-1}(R\cdot Y_{incl-h}-Y_{decay-h})
\end{equation}
in which $R$ measures the number of inclusive photons divided by
the number of decay photons from all decay channels. $R$ is
independently measured by PHENIX~\cite{justin}.

%\begin{figure}[ht]
%\scalebox{0.3}
%{\includegraphics{fig/ppdirect.eps}\includegraphics{fig/auaudirect.eps}}
%\caption{Direct photon hadron per trigger yield in $\pp$ (left
%panel) and $\Au$ (right panel).}\label{fig:ppdirect}
%\end{figure}

\begin{figure}[ht]
\includegraphics[width=0.55\textwidth]{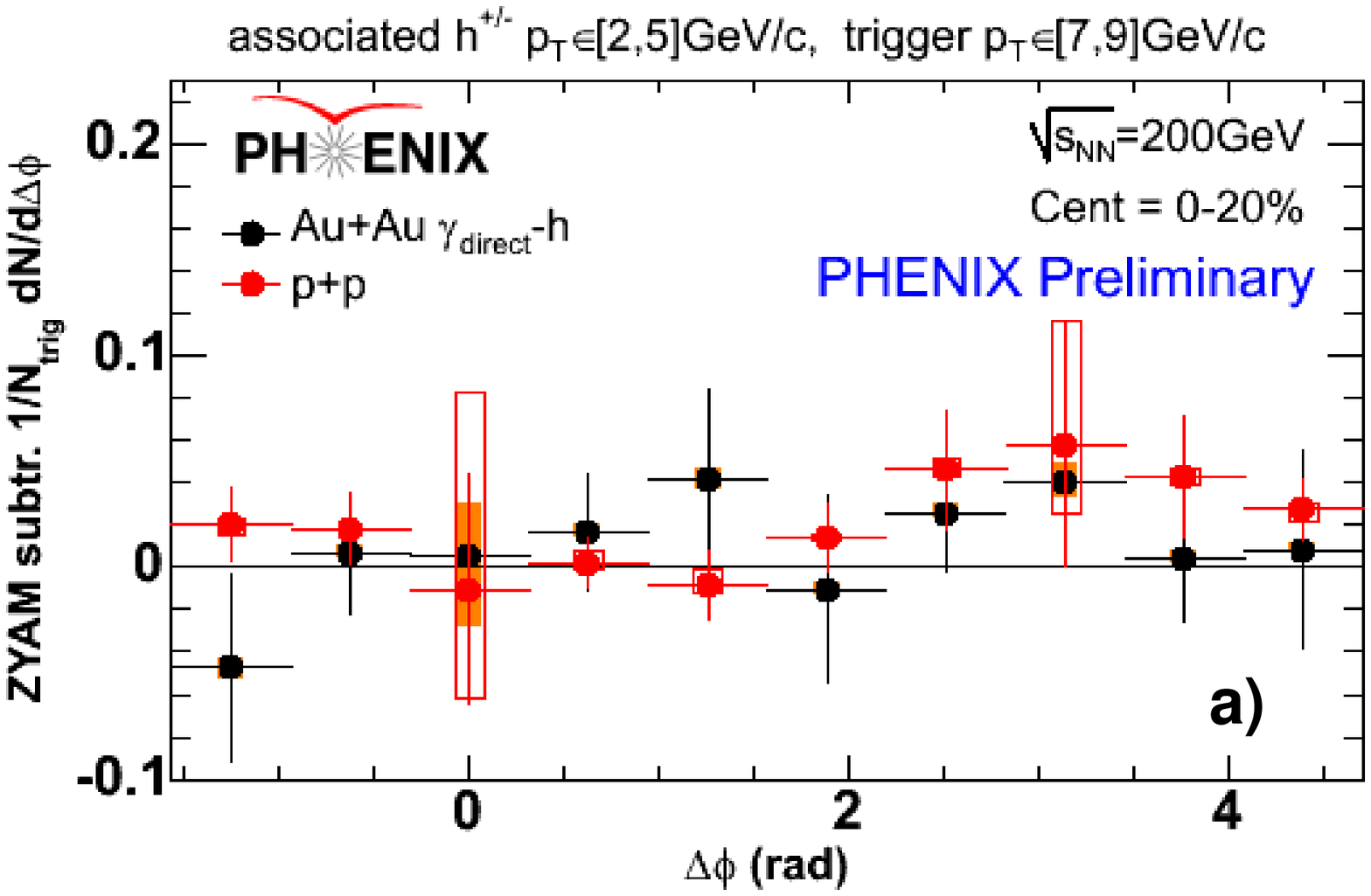}\hfill\includegraphics[width=0.45\textwidth]{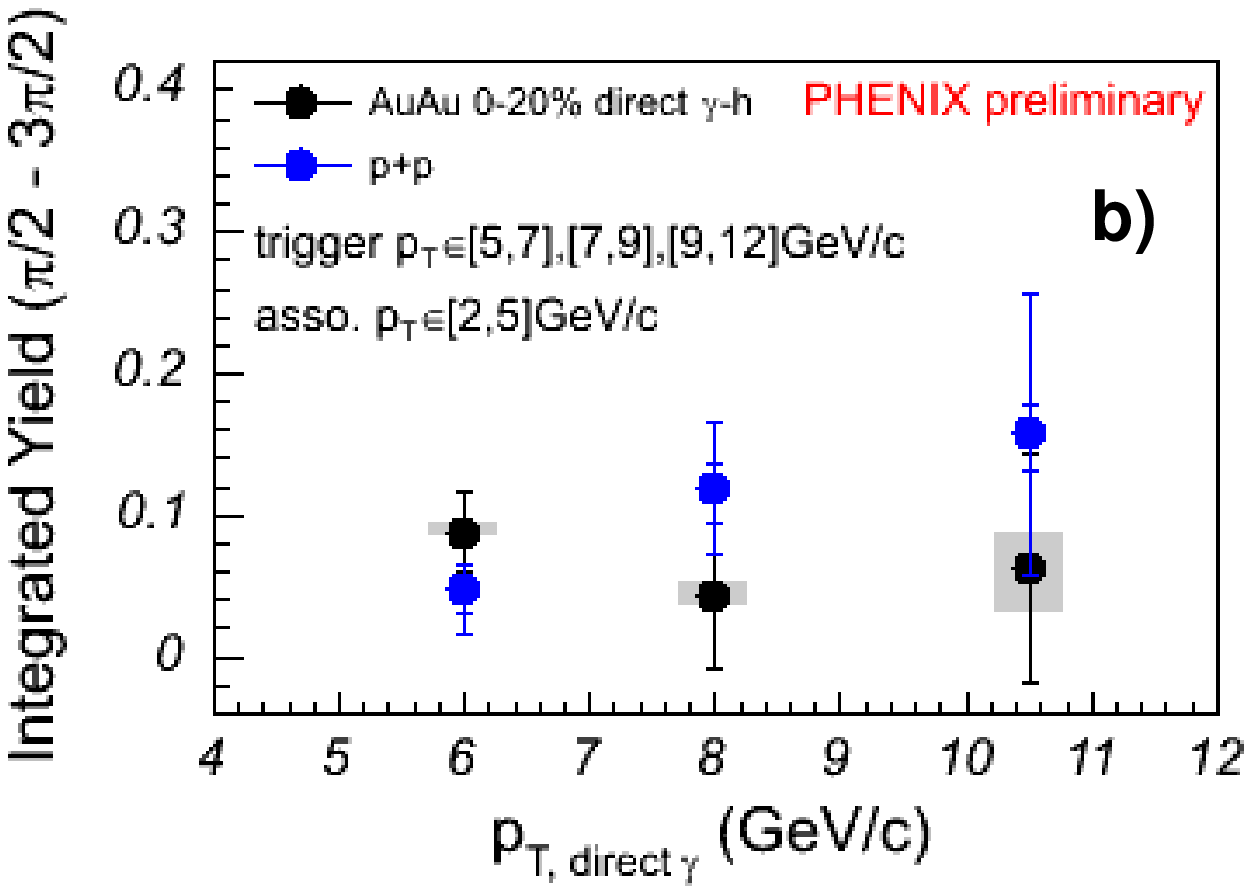}
\caption{(a) The direct photon-hadron per-trigger yield in $\pp$
(red) and $\Au$ (black) and (b) The integrated away side yields as
a function of trigger photon $p_T$ in $\pp$ (blue) and $\Au$
(black).}\label{fig:yieldcompare}
\end{figure}

The direct photon-hadron yield in $\pp$ is an important baseline
measurement. Fig.~\ref{fig:yieldcompare}(a) shows direct
photon-hadron per-trigger yield in $\pp$, a near-side yield
consistent with zero is seen, which is consistent with what one
would expect from direct photon and small fragmentation photon
contribution. A comparison between the $\pp$ per-trigger yield and
the PYTHIA simulation results shows a qualitative agreement. In
$\Au$ collisions, the direct photon-hadron per-trigger yield is
shown in Fig.~\ref{fig:yieldcompare}(a). The near-side yield is
consistent with zero and the away-side yield is small.

Comparing $\pp$ to $\Au$ in Fig.~\ref{fig:yieldcompare}(a), we see
some indication of away-side suppression. Although error bars are
large, there is a systematic trend that the $\pp$ yield is higher
than $\Au$. To make a quantitative statement, the away-side yield
is integrated over the $[\pi/2,3\pi/2]$ region.
Fig.~\ref{fig:yieldcompare}(b) shows the integrated away-side
yields as a function of trigger photon $p_T$. We observe an
increasing trend of yields in $\pp$, whereas yields in $\Au$ are
suppressed, especially when $p_T>7GeV/c$.

\section{Conclusions}

PHENIX has made precision measurements of high-$p_T$ dijets. We
observe near-side jet modification at large $p_{out}$. Away-side
yield is suppressed, whereas the width is unchanged. Also, the
away-side $I_{AA}$ is quantitatively consistent with $R_{AA}$.
Moreover, PHENIX has made the first measurement of high-$p_T$
direct photon-hadron yields. The near-side yield is consistent
with zero and the away-side yield is suppressed compared to yields
in $\pp$. Therefore these data indicate the modification of the
away-side jets from photon triggers in $\Au$.

\end{document}